\documentclass[journal=jctcce,article]{achemso}

\usepackage{amssymb}
\usepackage{amsmath}
\usepackage{amsfonts}
\usepackage{graphicx}
\usepackage{longtable}

\newcommand{\rcite}[1]{Ref.~\citenum{#1}}
\newcommand{\rcites}[1]{Refs.~\citenum{#1}}

\newcommand{\sref}[1]{Sec.~\ref{#1}}

\newcommand{\floor}[1]{\lfloor{#1}\rfloor}
\newcommand{\ceil}[1]{\lceil{#1}\rceil}

\newcommand{\ten}[1]{\mathsf{#1}}
\renewcommand{\vec}[1]{\boldsymbol{#1}}

\newcommand{\beq}{\begin{eqnarray}}
\newcommand{\eeq}{\end{eqnarray}}
\renewcommand{\d}{{\text{d}}}
\newcommand{\tr}{{{\text{Tr}}}}

\newcommand{\erf}{{{\text{erf}}}}

\newcommand{\LR}{\text{LR}}
\newcommand{\at}{\text{at}}

\newcommand{\half}{\frac{1}{2}}

\newcommand{\pr}{^{\prime}}

\newcommand{\vr}{\vec{r}}

\newcommand{\vrp}{\vec{r}\pr}

\newcommand{\degree}{\ensuremath{^{\circ}}}

\usepackage{color,xcolor}
\usepackage[normalem]{ulem}

\newcommand\TG[2]{\textcolor{green}{[TG] \sout {#1} {#2}}}

\newcommand\Note[2]{\textcolor{cyan}{[[ Note - #1 ]] {#2} [[ End note ]]}}

\newcommand{\comment}[1]{}

\def\nocomments{1}

\ifdefined\shortcomments

  \renewcommand\TG[2]{\textcolor{green}{#2}}
  
  \renewcommand\Note[2]{}
\fi

\ifdefined\nocomments

  \renewcommand\TG[2]{#2}
  
  \renewcommand\Note[2]{}
\fi

\title{A fractionally ionic approach to polarizability and
van der Waals many-body dispersion calculations}
\author{Tim Gould}
\email{t.gould@griffith.edu.au}
\affiliation{Qld Micro- and Nanotechnology Centre, %
Griffith University, Nathan, Qld 4111, Australia}
\author{S\'ebastien~Leb\`egue}
\email{sebastien.lebegue@univ-lorraine.fr}
\affiliation{Universit\'e de Lorraine,  Vand{\oe}uvre-l\`es-Nancy, F-54506, France}
\affiliation{CNRS, CRM2, UMR 7036, Vand{\oe}uvre-l\`es-Nancy, F-54506, France}
\author{J\'anos~G.~\'Angy\'an}
\email{janos.angyan@univ-lorraine.fr}
\affiliation{Universit\'e de Lorraine,  Vand{\oe}uvre-l\`es-Nancy, F-54506, France}
\affiliation{CNRS, CRM2, UMR 7036, Vand{\oe}uvre-l\`es-Nancy, F-54506, France}
\affiliation{Department of General and Inorganic Chemistry, Pannon University, Veszpr\'em, H-8201, HUNGARY} 
\author{{Tom\'a\v{s} Bu\v{c}ko}}
\email{bucko@fns.uniba.sk}
\affiliation{Department of Physical and Theoretical Chemistry, Faculty of Natural Sciences, Comenius University in Bratislava, Mlynsk\'{a} Dolina, Ilkovi\v{c}ova 6, SK-84215 Bratislava, Slovakia}
\affiliation{Institute of Inorganic Chemistry, Slovak Academy of Sciences,
D\'ubravsk\'a cesta 9, SK-84236 Bratislava, Slovakia} 

\begin{document}

\begin{abstract}
  By explicitly including \emph{fractionally ionic} contributions
  to the polarizability of a many-component system we are able to
  significantly improve on previous atom-wise many-body van der Waals
  approaches with essentially no extra numerical cost.
  For non-ionic systems our method is comparable in accuracy to
  existing approaches. However, it offers substantial improvements 
  in ionic solids, e.g. producing better
  polarizabilities by over 65\% in some cases.
  It has particular benefits for two-dimensional
  transition metal dichalcogenides,
  and interactions of H$_2$ with modified coronenes -
  ionic systems of nanotechnological interest.
  It thus offers an efficient improvement on existing approaches,
  valid for a wide range of systems.
\end{abstract}

\section{Introduction}

Over the past decade there has been a resurgence of interest
within the electronic structure community in the inclusion of
van der Waals (vdW) dispersion forces in \emph{ab initio}
calculations. As a result, a plethora of new approaches has
been developed\cite{DobsonDinte,
Lundqvist1995,ALL,Dion2004,Langreth2005,
Becke2005-XDM,
Grimme2004,Grimme2006,Grimme2010,
Tkatchenko2009,Tkatchenko2012,Dobson2012-JPCM}
that allow van der Waals forces (more generally implemented as
potentials)
to be included alongside conventional density functional
approximations\cite{Becke1988,GGA,LYP}.
Recent theoretical and experimental progress on
``non-additivity'' of van der Waals
interactions\cite{Dobson2006,Gould2008,Gould2009,Lebegue2010,%
DiStasio2012,Gobre2013,Dobson2009,Dobson2012-JPCM,Dobson2014-IJQC,%
Gould2013-Cones,Tao2014-Surface,Dobson2014-1,%
Tkatchenko2015-Review,Reilly2015,%
Ambrosetti2016,Tao2016-fullerene,Dobson2016-LRT} has also seen
a more fundamental shift in thinking about van der Waals
forces and the role they play in nanomaterials.

This reinvigorated interest in vdW forces has both
parallelled and been driven by increasing interest in
layered ``2D'' materials which are held together almost
exclusively by weak dispersion (vdW) forces. Without
accurate models of vdW forces it is very difficult to predict
even qualitatively correct properties of such layered systems
- including lattice parameters, binding energies and elastic moduli.
Thus the inclusion of vdW effects is vital for the
modeling of layered
structures; and layered systems provide an important
test of their accuracy\cite{Bjorkman2012-Review}.

Some of the more popular approaches for including vdW forces
are that of Tkatchenko and Scheffler\cite{Tkatchenko2009} (TS);
related many-body dispersion (TS-MBD)
schemes\cite{Tkatchenko2012,DiStasio2012,Ambrosetti:14,Kim:16};
and self-consistent screening\cite{Tkatchenko2012,Ambrosetti2016} (TS-SCS)
schemes. These approaches
allow vdW forces to be included in calculations
via a point-dipole model, with excellent numerical efficiency
and generally good qualitative results.
The methodology behind these approaches can be 
summarized as follows:
\begin{enumerate}
\item Pre-calculate $C_6$ coefficients and static dipole
  polarizabilities $\alpha_0$ for isolated atoms using
  high-level electronic structure theory.
\item  Approximate dipole polarizability properties of a
  molecule or material as a superposition of polarizabilities
  of free-atoms centered on their nuclei.
\item  Use the free-atom properties, appropriately rescaled
  by the effective volume of embedded atoms, to calculate a model
  frequency dependent (dynamic) polarizability (see Sec.S1 in the 
  Supporting Information\cite{SuppMat}).
\item  Use the model polarizability to calculate the dispersion
  energy. This is done either by summing over pairwise contributions,
  as in the original TS theory\cite{Tkatchenko2009},
  or by solving a many-body screening equation that includes
  non-additive terms, as in
  many-body approaches\cite{Tkatchenko2012,DiStasio2012,Ambrosetti:14}.
\end{enumerate}

The TS, TS-MBD and TS-SCS
schemes have shown great success in improving
the qualitative and quantitative accuracy of \emph{ab initio}
calculations of many 
systems\cite{Tkatchenko:09,Lilienfeld:10,AlSaidi:12,Bucko:13a}.
\TG{}{This reflects their ability to capture Type-A (Dobson-A) and
many-body Dobson-B effects, according to the classificiation
of Dobson\cite{Dobson2014-IJQC}.}
However, it was shown in \rcite{Bucko:13a}
that the TS and TS-SCS schemes may perform poorly for
ionic systems, due to their use of \emph{atomic}
rather than  \emph{ionic} volumes.
As a solution, an alternative to stage~3 was
proposed\cite{Bucko:13a,Bucko2014-Iterative}
that accounts for rescaled \TG{}{(via the ratio of the
effective embedded volume to the free atom volume)}
densities of \emph{ions},
rather than pure atoms. This was found
to improve results in many ionic systems.
However, while the improvements from the ionic rescaling go
some way to improving results for ionic materials, it still
fails to capture the full ionic physics of the systems,
leading e.g. to poor dipole polarizabilities of ionic systems.

In this work we will go further than
\rcite{Bucko2014-Iterative} and take into account
the effect of both \emph{ionic charge and volume} on the effective
static polarizabilities $\alpha_0$ and $C_6$ coefficients
of the system. More precisely, we will take into account
the effect of the ionic charge on the frequency-dependent
dipole polarizibilities, drawing from a recently published
ionic polarizability dataset produced by two of the
authors of the present work\cite{Gould2016-C6}.
Our new method thus modifies stages~1, 3 and 4 of TS-derived
schemes (\emph{vide supra}) by including:
a) ionic charge in the high-level calculations in stage~1,
via a new ionic polarizability dataset\cite{Gould2016-C6};
b) ionic (rather than atomic) volume scaling of the
polarizabilities in stage~3, and
c) a more accurate model of frequency-dependent polarizabilities
in stages~3 and~4. 

We will first develop a theory of embedded fractional ions by: i) showing how a dataset of \emph{integer} ions can be used to
generate \emph{fractional} ions;
ii) introducing a ``\TG{rescaling}{remapping}'' correction to the MBD
equations that corrects for rare, but sizeable, unphysical
interactions in some materials.
We will finally show that the explicit inclusion of fractional
charge in the point-dipole model substantially
improves results in various systems, especially in ionic ones.
Our new scheme is particularly
important for ionic layered transition metal dichalcogenide
(TMD) systems, like MoS$_2$ where other schemes perform very poorly.
In these systems the ionic scheme gives results
on a par with sophisticated \TG{}{electron structure calculations
using} random-phase approximation
(RPA)\cite{Lebegue2010,Eshuis2012,Dobson2012-JPCM,Bjorkman2012-Review}.

\section{Theory}

Our new scheme is closely related to the work of Tkatchenko
and co-workers\cite{Tkatchenko2009,Tkatchenko2012,DiStasio2012},
as improved by Bu\v{c}ko \emph{et al}\cite{Bucko2014-Iterative}.
Notably, it employs the same iterative Hirshfeld partitioning and
the same effective volume scaling as
\rcite{Bucko2014-Iterative}. However, it
\emph{differs crucially} through the polarizability model
it employs, because it incorporates fractionally ionic
effects and uses a significantly improved screening.
The new recipe thus becomes:
\begin{enumerate}
\item Pre-calculate dipole polarizabilities not only for 
free atoms, but also for non-interacting \emph{free ions}
  using high-level electronic structure theory.
\item Approximate the dipole polarizability properties of a
  molecule or a material via non-interacting \emph{free-ion}
  polarizabilities centered on the nuclei.
\item Use the free-ion polarizabilities to obtain
  \emph{fractional ionic (FI)} polarizabilities
  $\alpha_p^{\text{FI}}$,
  evaluated at the \emph{fractional} electron number $N_p$,
  and rescaled\footnote{We tested more sophisticated models of
    rescaling from \rcite{Gould2016-C6Vol} but found it did not
    noticeably improve results.}
  by the effective volume of the embedded charge, 
  to calculate the effective fractional ionic polarizabilities
  (see Sec.S1 in the 
  Supporting Information\cite{SuppMat} for further details):
  \begin{equation}\label{eqn:alphaTS2}
    \alpha_p^{\text{AIM}}(i\omega;N_p) = \alpha_p^{\text{FI}}(i\omega;N_p) 
    \frac{V^{\text{eff}}_p(N_p)}{V^{\text{FI}}_p(N_p)}.
  \end{equation}
\item Use the effective, fractional ionic embedded
  polarizabilities distributed on
  the atomic sites to calculate the dispersion
  energy, either using the additive or many-body method
  with eigenvalue \TG{rescaling}{remapping} (discussed later).
\end{enumerate}
These items are treated in detail over the following
sections. We first introduce the fractional ionic  scheme
[\sref{ssec:Frac}];
we then discuss the polarizability model used in the
fractional ionic scheme [\sref{ssec:Model}];
and we finally explain a ``\TG{rescaling}{remapping}'' scheme
that ensures that the energy and forces
behave reasonably even for some of the systems where the original scheme
fails [\sref{ssec:Rescale}].

\subsection{Properties of fractional ions}
\label{ssec:Frac}

Before introducing the details of the new method, we should first
discuss here what is meant by ``fractional ions''. In an embedded
system this concept is fairly well defined,
at least within partitioning schemes 
such as Bader~\cite{Bader:book}, Hirshfeld~\cite{Hirshfeld:77},
or iterative Hirshfeld~\cite{Bultinck:07} atoms-in-molecules (AIM)
approaches: the ``fractional ion'' is defined as a
sub-system of a larger molecule or material with an integral nuclear
charge $Z$ and a non-integer number of electrons $N$, whose
overall charge $f=Z-N$ is thus non-integral in general. 
This is related to a representative free
fractional ion with equal non-integer charge.
The use of  fractional ions raises interesting
fundamental questions, whilst offering significant practical
improvements. We thus dedicate the following part of this Section
to discuss the nature and practical consequences of
using such fractional ions.

Any isolated system with non-integer electron number must be
treated via a quantum state ensemble composed
of wave functions with different integer electron numbers.
This applies at the fully interacting and density functional
theory levels\cite{Perdew1982}.
In an isolated \emph{atomic-like} system, such as the interaction-free fractional ions
that we will use, these states will all share the same form of nuclear
potential $-Z/r$. Furthermore, at low temperatures, the ensemble
of non-interacting free ions with non-integer $N$ electrons will be formed
only from states with $\floor{N}=M$ and $\ceil{N}=M+1$ electrons
(i.e.\ the largest integer smaller and the smallest integer larger
than $N$).
As a consequence, many ground state properties of
such systems obey the well-established piecewise linear
relationship\cite{Perdew1982,Ayers2008}
\begin{align}
  P_N=&(N-M)P_{M+1} + (M+1-N)P_{M}
  \nonumber\\
  \equiv& fP_{M+1} + (1-f)P_M.
  \label{eqn:piecewise}
\end{align}
Examples include the energy $E_N$, electron density $n_N(\vr)$, and
atomic kinetic energy $T_N$\cite{Levy2014}.

The $C_6$ coefficient is not one of those properties.
However, we will argue below that the dipole polarizability
at non-integer $N$ \emph{is} piecewise linear and thus
does obey Eq.~\eqref{eqn:piecewise}.
If we then assume that the usual Casimir-Polder formula
\begin{align}
  C_{6,AB}=\frac{3}{\pi}\int d\omega
  \alpha_{Z_A,N_A}(i\omega)\alpha_{Z_B,N_B}(i\omega)
  \label{eqn:CP}
\end{align}
holds for fractional ions, this gives us a recipe for calculating
$C_6$ coefficients in terms of the imaginary frequency polarizabilities
\begin{align}
  \alpha_{Z,N}(i\omega)=&\int\d\vr\int\d\vr' xx'\chi_{N}[v_Z](\vr,\vrp;i\omega).
  \label{eqn:alpha}
\end{align}
Here $\chi_{N}[v_Z](\vr,\vrp;i\omega)$ is the density response of
an $N$ electron atom (where $N$ can be non-integer)
in potential $v_Z=-Z/r$.

Our argument is as follows:
By definition of the density response
$\chi=\frac{\delta n}{\delta v}$, we know that
$N$ electrons (kept fixed) in a potential
$v_Z + \delta v e^{\omega t}$ with $\delta v\to 0$ have a density
\begin{align}
  n'(\vr) =& n(\vr) + \int\d\vrp\chi(\vr,\vrp) \delta v(\vrp) e^{\omega t}
\end{align}
where $n(\vr)$ is the density with $\delta v=0$. In the case of a
non-integer electron number ensemble we further know that the
density
\begin{align}
  n_{N}(\vr)=&fn_{M+1}(\vr) + (1-f)n_M(\vr)
  \label{eqn:nN}
\end{align}
is piecewise linear between the electron densities of the adjacent
integer systems.

But Eq.~\eqref{eqn:nN} must hold also for any infinitesimally small
change (at least at zero frequency, but most likely at
finite frequency too\cite{Pribram-Jones2016-Thermal}) to the potential,
provided we assume that symmetries are preserved or are correctly
accounted for in any approximations employed\footnote{Note that
  our aim here is not to provide a rigorous proof but to provide
  a justification for the interpolation in practice}.
Thus we see also
\begin{align}
  n'_{N}(\vr)=fn'_{M+1}(\vr) + (1-f)n'_{M}(\vr).
\end{align}
And if $n'$ is piecewise linear and $n$ is piecewise linear,
it directly follows that the density response
\begin{align}
  \chi_{N}(i\omega)=&f \chi_{M+1}(i\omega) + (1-f)\chi_{M}(i\omega)
\end{align}
must be piecewise linear too, since $n'=n+\int \chi\delta v$.
Finally, using Eq.~\eqref{eqn:alpha}, we see that
\begin{align}
  \alpha_{Z,N}(i\omega)=&f \alpha_{Z,M+1}(i\omega) + (1-f)\alpha_{Z,M}(i\omega).
\end{align}
That is, the polarizability is piecewise linear in the electron number.

This means that we only need to calculate polarizabilities for
integer numbers of electrons, and interpolate to find the remaining
fractional ionic values. This ability to interpolate from integer
cases to non-integer cases is crucial for the success of our method as
it keeps the amount of external data sourced from high-level
calculations at a manageable size.

\subsection{Polarizability model}
\label{ssec:Model}

In the original work of Tkatchenko and Scheffler~\cite{Tkatchenko2012}, 
a simple relationship
\begin{align}
  \alpha(i\omega)=&\frac{\alpha(0)}{1 + \omega^2/\eta^2},
  &
  \eta=&\frac{4C_6}{3\alpha(0)^2}
  \label{eqn:TSPolar}
\end{align}
was assumed between the frequency dependent dipole polarizability,
the $C_6$ coefficient and the static dipole polarizability of the
isolated atom. This approximation is based on a one-pole
model of the polarizability of an isolated atom, with $\eta$
set consistent with Eq.~\eqref{eqn:CP}.

In the present work we can no longer rely on Eq.~\eqref{eqn:TSPolar}
due to the use of linear combinations of polarizabilities,
each with a different value of $\eta$.
Rather than generalizing Eq.~\eqref{eqn:TSPolar}, we
instead use the two-pole parameterizations of
frequency dependent atomic and ionic polarizabilities
recently published by two of us\cite{Gould2016-C6}.
These models are then combined, as per
the piecewise linear formula, into a four-pole model for the
fractional ions. This allows us
to improve the accuracy of the integer atomic polarizabilities
and ensure that our fractional ions are treated
correctly.

For each atom and ion with integer electron number $M$
we thus employ a polarizability model
\begin{align}
  \alpha_{Z,M}(i\omega)=&\sum_{c=1,2} \frac{a_c(Z,M)}{\omega^2 + \Omega_c(Z,M)^2}
\end{align}
involving two Lorentzian functions.
Then, using the piecewise linearity of $\alpha$, we extend
this to \TG{}{arbitrary non-integer electron number $N$ using}
\begin{align}
  \alpha_{Z,N}(i\omega)=&
  f\sum_{c=1,2} \frac{a_c(Z,M+1)}{\omega^2 + \Omega_c(Z,M+1)^2}
  \nonumber\\&
  + (1-f)\sum_{c=1,2} \frac{a_c(Z,M)}{\omega^2 + \Omega_c(Z,M)^2}.
  \label{eqn:alphaFrac}
\end{align}
\TG{}{Here we use the properties of systems with neighbouring
integer electron numbers $M=\floor{N}$ and $M+1$, and then
linearly interpolate using the remaining fractional charge $f=N-M$}.
The final model thus employs four Lorentzian functions. The values 
of parameters $a_c(Z,M)$ and $\Omega_c(Z,M)$ used in this work are 
available for neutral atoms 
and ions of elements from the first six rows of Periodic Table in the 
Supporting Information of \rcite{Gould2016-C6}.

Finally, we can perform all integrals analytically in Eq.~\eqref{eqn:CP}
to derive the $C_6$ coefficient (if needed) between two fractionally
charged ions $A$ and $B$ in terms of our pretabulated coefficients.
Alternatively, in the case of the many-body theory the direct
calculation of $C_6$ is unnecessary, and instead we can use
Eq.~\eqref{eqn:alphaFrac} directly.

\subsubsection{A note on embedded ions}

Free-standing ions can be very different to ions embedded
in a larger system such as a molecule or a material. For example,
the surrounding environment of an anion has a substantial effect on its
outermost electron(s). They are only very weakly bound in the
free-standing case, with an asymptotic effective potential
going to zero as $~r^{-3}$. When embedded, the
presence of other electrons and nuclei should be represented by
an effective confining potential, which can be
approximated by a positive power law function of $r$ such as
$(r/r_a)^{\sigma_a}$ where $r_a$ is an effective embedding radius
and $\sigma_a$ governs the sharpness of the embedding.

The same electrons that are the most sensitive to the embedding
environment, namely electrons in the outermost electronic shell(s),
are the ones that contribute the most to the polarizability.
Thus the polarizability of an embedded system is highly sensitive
to its environment and care must be taken in considering what
``ions'' we use in our high-level fractional ionic calculations.
This is particularly pertinent for open shell systems which are
likely to be the most sensitive to the environment.

To this end we use the model polarizabilities from the
``minimal chemistry'' database of \rcite{Gould2016-C6}.
This provides \emph{ab initio} dynamic polarizabilities
for the first 6 rows of the periodic table that are
specifically tailored to provide  chemically plausible
free anions, calculated as a combination of
ensemble DFT\cite{Gould2012-LEXX}
and time-dependent DFT\cite{Gould2012-RXH,Gould2013-Aff}.
It also allows for the evaluation of double anions, which
are problematic in a self-consistent scheme.
In short, self-consistent results are included
for neutral atoms, cations, and closed shell monoanions;
while non self-consistent data (based on the relevant neutral atom)
is provided for the remaining anions.


\subsection{Eigenvalue remapping}
\label{ssec:Rescale}

Variants of the MBD method involve (see e.g. discussion
in \rcites{DiStasio2012,Bucko:16})
the solution of the screening equation
\begin{align}
  E_{\text{disp}}=&-\int_0^{\infty}\frac{d\omega}{2\pi}
  \tr\{\ln[\ten{1}-\ten{A}_{\LR}(\omega)\ten{T}_{\LR}]\}
  \label{eqn:EdispATLR}
\end{align}
to obtain dispersion energies.
Here $\ten{A}_{\LR}$ is an atom-wise model of the dipole polarizability
in the system which differs only in some details between
different schemes (like MBD@rsSCS, and the 
new MBD@rsSCS/FI model introduced here), and
$\ten{T}_{\LR}$ is the (long-range) dipole interaction tensor.
Calculations thus involve a tractable $3N_{\at}\times 3N_{\at}$ matrix
equation, where $N_{\at}$ is the number of atoms/ions in the
system. In the case of a periodic system $N_{\at}$ is the number of
atoms/ions in the unit cell and we must also include a
sum over $\vec{k}$ points in the irreducible Brillouin zone.

Full details of TS-MBD variants, including the generalization
of Eq.~\eqref{eqn:EdispATLR} to periodic systems, can be found
in \rcite{Bucko:16}.
In summary, the matrix used in the MBD@rsSCS scheme is defined as
$[A_{\LR}(\omega)]_{pq \alpha\beta}=\tilde{\alpha}_{p}(i\omega)
\delta_{pq}\delta_{\alpha\beta}$
where $\tilde{\alpha}_{p}(i\omega)$ is the frequency-dependent
polarizability of the embedded atom or ion (here $p$ and $q$ refer
to atom indices and $\alpha$ and $\beta$ refer to Cartesian indices
$x$, $y$ and $z$) obtained by solving the short-range screening 
equation 
\begin{equation}
  \tilde{\alpha}_p(i\omega) = \alpha_p(i\omega)
  \left ( 1-  \sum_q T_{SR,pq}(i\omega) \, \tilde{\alpha}_q(i\omega) \right ). 
  \label{eqn:SCS}
\end{equation}
Here $T_{SR}$ is a short-range dipole-dipole
interaction tensor.
\TG{}{$\tilde{\alpha}(i\omega)$ used to define $A_{LR}$ is one
third of trace of the matrix solution of \eqref{eqn:SCS}.}

In conventional MBD@rsSCS 
the term $\alpha_p(i\omega)$ in Eq.~\eqref{eqn:SCS}
comes from Eq.~\eqref{eqn:TSPolar}.
In our new FI scheme, $\alpha_p(i\omega)$ is replaced by
$\alpha_{Z_p,N_p}(i\omega)$ defined
for fractional ions using Eq.~\eqref{eqn:alphaFrac}. The replacement
of Eq.~\eqref{eqn:TSPolar} by Eq.~\eqref{eqn:alphaFrac} thus 
includes embedded ionic charge explicitly
[through local electron number $N_P$ in Eq.~\eqref{eqn:alphaFrac}]
and implicitly [via the volume scaling in Eq.~\eqref{eqn:alphaTS2}].
The matrix $[T]_{\LR,pq \alpha\beta}$ is used identically in both
methods, and is related to the second spatial derivative
of a modified Coulomb potential.
We refer the reader to Section~II of \rcite{Bucko:16} for furhter
details.

Once we have $\ten{A}_{\LR}$ and $\ten{T}_{\LR}$ we can solve 
Eq.~\eqref{eqn:EdispATLR}
using the eigenvalues $x_n$ of the Hermitian matrix
$\ten{X}(\omega)=-\ten{T}_{\LR}^{\half}\ten{A}_{\LR}(\omega)\ten{T}_{\LR}^{\half}$%
\footnote{Note that identical results are obtained using
$\ten{A}_{\LR}(\omega)\ten{T}_{\LR}$ directly. Indeed our calculations
use this direct form. For the manuscript we use the Hermitian form
to simplify working.}.
Here the equalities
$\tr[f(-\ten{A}\ten{T})]=\tr[f(\ten{X})]=\sum_n f(x_n)$ give
\begin{align}
  E_{\text{disp}}=&-\int_0^{\infty}\frac{d\omega}{2\pi}
  \sum_n\ln[1+x_n(\omega)].
  \label{eqn:EdispEV}
\end{align}
The eigenvalues $x_n(\omega)$ should obey $x_n>-1$
and $\sum_nx_n=0$ due to constraints and sum rules
imposed by the physics of interacting systems.

Unfortunately certain systems, including transition-metal dichalcogenides
(discussed in more detail later), highlight a physical and numerical
deficiency in the MBD approach that manifests itself most obviously
in Eq.~\eqref{eqn:EdispEV}. In these systems long-range
and low-frequency dipole coupling can lead to some values of $x_n$
becoming less than $-1$ due to inconsistencies in the \emph{model}
polarizability and screening equations. This problem is related to the
well-known phenomenon of
``polarization catastrophe''~\cite{Tosi:67,Faux:71} \TG{}{and occurs
when atoms are too close together - rendering the Taylor expansion
of the Coulomb potential unstable}.

To solve the system of self-consistent screening equations in these 
pathological cases we need to fix these unphysical eigenvalues. Given the
success of MBD in many systems, this fix must maintain its good
properties in typical cases, but improve on it in the pathological cases.
Our solution involves ``\TG{rescaling}{remapping}''
the eigenvalues of the system
to avoid unphysical cases whilst maintaining continuity in energy and forces
with respect to changes in the geometry. Specifically, we
replace $\sum_n\ln(1+x_n)$ by
$\sum_n\{\ln(1+\tilde{x}_n)-\tilde{x}_n\}$ where
\begin{align}
  \tilde{x}_n=&\begin{cases}
    x_n & x_n\geq 0, \\
    -\erf\big[(\frac{\sqrt{\pi}}{2}|x_n|)^4\big]^{1/4} & x_n<0.
  \end{cases}
  \label{eqn:EReqn}
\end{align}
The additional term  in the sum ($-\sum_n\{\tilde{x}_n\}$)
is required to correct
$\tr[\tilde{\ten{X}}]$ which, unlike $\tr[\ten{X}]$, is not
guaranteed to be zero.

The \TG{rescaled}{remapped} eigenvalues $\tilde{x}_n$
obey $\tilde{x}_n> -1$ for finite $x_n$
so that $\ln(1+\tilde{x}_n)$ is well-defined in all cases.
Furthermore, $\tilde{x}_n$ is a continuous function of
$x_n$ with smooth \TG{first and second}{leading} derivatives. It thus
produces meaningful \emph{and} smooth energies and forces
for all reasonable geometries.
\TG{}{The specific form of the \emph{ad hoc} mapping
from $x$ to $\tilde{x}$
is somewhat arbitrary. Equation~\eqref{eqn:EReqn} was chosen as it
straightforwardly ensures that: a) $\tilde{x}>-1$ as required,
b) $\tilde{x}=x + O(x^9)$ as $x\to 0$, leading to
correct derivatives to all interesting orders and
c) $|\tilde{x}-x|<0.01$ for $x>-0.8$, meaning the correction is only
significant in extreme cases.}

This \TG{rescaling}{eigenvalue remapping}
is mathematically equivalent to transformation
of the model polarizability $\ten{A}_{\LR}$, albeit in a fairly
non-conventional fashion. The final energy expression can thus
equivalently be written as
\begin{align}
  E_{\text{disp}}=&-\int_0^{\infty}\frac{d\omega}{2\pi}
  \sum_n\{\ln[1+\tilde{x}_n(\omega)]-\tilde{x}_n\}
  \\
  \equiv &-\int_0^{\infty}\frac{d\omega}{2\pi}
  \tr\{\ln[\ten{1}-\tilde{\ten{A}}_{\LR}\ten{T}_{\LR}]
  -\tilde{\ten{A}}_{\LR}\ten{T}_{\LR}\}
\end{align}
where $\tilde{\ten{A}}_{\LR}(\omega)
=-\ten{T}_{\LR}^{-\half}[\sum_n\tilde{x}_n(\omega)\ten{V}_n(\omega)]
\ten{T}_{\LR}^{-\half}$.
Here $[V_n]_{ij}=v^*_{n,i}v_{n,j}$ for $\vec{v}_n$ the eigenvector
of $\ten{X}$ with eigenvalue $x_n$.

\TG{rescaling}{Eigenvalue remapping}
is an integral part of our new FI scheme, as it
ensures that the effective polarizability tensor obeys the
necessary constraints. This leads to improved dispersion physics
and consequentially better energies when
\TG{rescaling}{remapping} is needed,
and almost no change when it is not.

We note that eigenvalue \TG{rescaling}{remapping} can 
be optionally linked with the original MBD@rsSCS scheme.
This variant, denoted by MBD@rsSCS+ER, is able to fix 
the pathological behavior of the MBD@rsSCS method in certain, 
albeit in not all, cases. However, when the solution exists,
the  results
obtained by MBD@rsSCS+ER are generally poorer than those obtained
by the new  MBD@rsSCS/FI method. Some numerical MBD@rsSCS+ER
results are presented in the
Supporting Information\cite{SuppMat}.
Performance for ``well-behaved'' systems is affected 
by eigenvalue \TG{rescaling}{remapping} only marginally 
(e.g. Sec.~\ref{ssec:molbench}) in both the MBD@rsSCS+ER
and MBD@rsSCS/FI approaches.

\section{Numerical results}

We examine the usefulness of the proposed method 
in a number of practical applications. First, we will 
demonstrate that the dipole polarizabilities 
computed using fractional charges (Eq.~\eqref{eqn:alphaFrac})
outperforms the 
approach based on neutral atoms. Second, 
we combine the frequency dependent dipole polarizibilities  
with the many-body dispersion correction method 
of Ambrosetti et al.~\cite{Ambrosetti:14}
(PBE+MBD@rsSCS) and show that this new version 
of the method performs significantly better, especially for 
for systems with strong ionic behavior. 

\TG{}{All calculations were carried out with the periodic DFT code 
VASP~5.4.1\cite{Kresse:93b,Kresse:96,Kresse:99},
modified to include the FI approach. The FI approach will be
available in the next official release of VASP.
A patch is presently available by request to the authors.
Setting keyword {\tt IVDW=263} activates the FI treatment.
For the sake of completeness we note that
the combination of keywords {\tt IVDW=202} and {\tt ITIM=1} activates
conventional MBD\cite{Bucko:16} with the eigenvalue remapping,
albeit we do not recommend this approach for routine applications.}

\TG{The DFT part of calculations have been performed using the}%
{For the semi-local approximation we use the}
Perdew-Burke-Ernzerhof (PBE) functional~\cite{Perdew:96}.
The electronic energies were converged to
\TG{}{self-consistency with} an accuracy of $10^{-7}$ eV.
The $\vec{k}$-point\TG{s}{} grids and 
plane-wave cutoffs used in calculations are summarized in 
Tab.~\ref{tab:setting} and Tab.S5 in the 
Supporting Information\cite{SuppMat}.  
Drawings of structures presented in this work 
have been created using the program
VESTA~\cite{Momma:11}.

As with conventional TS and \TG{}{its} variants, the choice of
semi-local approximation is somewhat flexible.
For this work we focus on PBE as it allows ready
comparison with past calculations. However, the
method presented here can, in principle, be coupled with
alternative approximations like \TG{popular}{accurate}
meta-GGAs\cite{TPSS,SCAN,Tao2016-MGGA}.
In such cases the adjustable parameter $\beta$ would need to
be recalculated to properly account for the improved
short-range interaction in meta-GGA functionals.

\subsection{Polarizabilities of atoms in cubic ionic crystals}
On the basis of experimental data, polarizabilities of 
cubic crystals can be obtained from the Clausius-Mossotti equation
\begin{equation}
  \alpha_m = \frac{3V}{4\pi}
  \left(\frac{\epsilon_{\infty}-1}{\epsilon_{\infty}+2}\right ),
\end{equation}
where $V$ is the volume per formula unit (fu)
and $\epsilon_{\infty}$ 
is the measured high-frequency dielectric constant. In this work we use
experimental results for  from \rcites{Fowler:85,vanVechten:69}
to compute $\alpha_m$ for a set of 18 cubic crystals that ranges from
6.2 au (LiF) to 61.9 au (CsI). 

Using the static AIM polarizabilities, 
polarizability of atoms in crystals are computed by
solving the  self-consistent screening (SCS) equation 
described in the Supporting Information\cite{SuppMat}.
The following three models to determine AIM polarizabilities 
(differing in the type of reference object (i.e. atom or fractional ion) 
and partitioning scheme used)
have been considered: rescaling of polarizabilities 
of neutral atoms using \TG{}{effective volumes obtained from}
Hirshfeld partitioning (TS), 
rescaling polarizabilities 
of neutral atoms using \TG{}{volumes from} iterative Hirshfeld partitioning
(HI), and rescaling polarizabilities 
of fractional ions using \TG{}{volumes from}
iterative Hirshfeld partitioning (FI).

\begin{figure}[h]
  \caption{
    Deviations of theoretical polarizabilities (au/fu)
    for selected crystalline materials from their experimental 
    counterparts. The theoretical values have been obtained using:
    the model based on neutral atoms and Hirshfeld 
    partitioning (TS - red stars);
    the model based on neutral atoms and
    iterative Hirshfeld partitioning (HI - green triangles);
    and the new model based
    on fractional ions and iterative Hirshfeld partitioning
    (FI - blue circles).
    The thick solid black line indicates a perfect agreement between
    theory and experiment. The experimental results
    are from \rcites{Fowler:85,vanVechten:69}.
  }\label{fig:inCrystCompar}
  \includegraphics[width=.99\columnwidth]{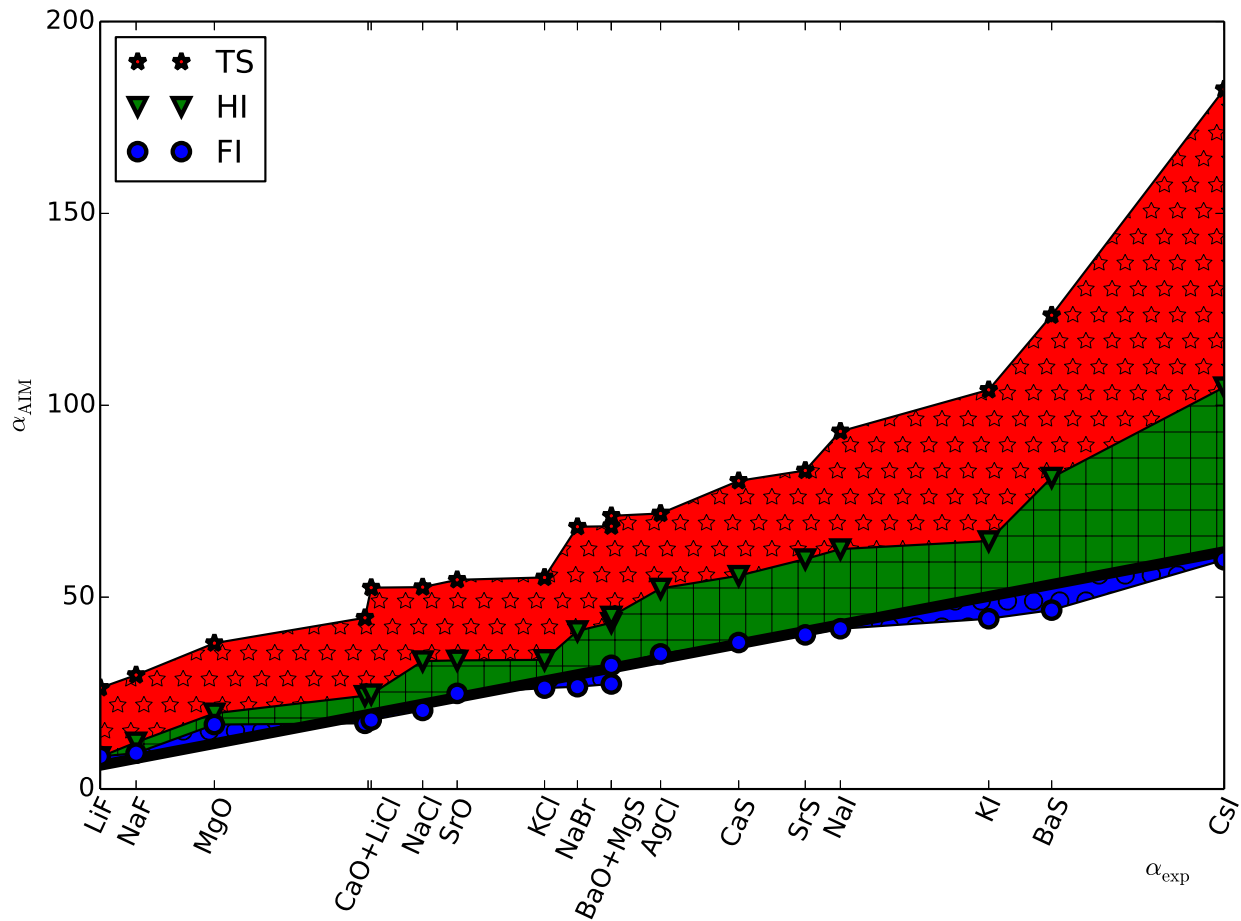}
\end{figure}

The TS model is used in the MBD@rsSCS scheme of
Ambrosetti~\cite{Ambrosetti:14},
while the FI model is the basis of our new scheme. 
The polarizabilities computed using all three models are compared with 
experimental data in Figure~\ref{fig:inCrystCompar}, the numerical 
values for all systems are compiled in  Tab.~S1 in the
Supporting Information\cite{SuppMat}.
The TS model
strongly overestimates polarizabilities for all systems:
mean absolute  error (MAE) and mean absolute relative error (MARE)
being as large as 
$41.5$~au/fu (where again fu stands for formula unit)
and $157.4\%$, respectively. The use of iterative 
Hirshfeld partitioning leads to significant improvement but 
the error still remains very large (MAE=13.7 au/fu, MARE$=43.5\%$).
Finally, replacing neutral atoms as reference
by fractional ions (FI) leads to a further significant
reduction of error with respect to experiment
(MAE=6.5 au/fu, MARE$=22.9\%$). 

These results also highlight the importance of a full fractional ionic
treatment for e.g. self assembly or surface physics problems.
The interaction of atoms, layers and molecules with bulks
is highly dependent on the crystal polarizability. Here the
FI treatment gives a twofold improvement over the HI method and
a sixfold improvement over standard TS.

\subsection{Molecular systems}
\subsubsection{Benchmark sets S66$\times$8 and X40}\label{ssec:molbench}
The benchmark set S66$\times$8 of
\v{R}ez\'a\v{c} et al.~\cite{Rezac:11}
consists of configurations from dissociation curves for 66 molecular dimers.
For each dimer, eight points differing in the  distance
between monomers
ranging between 0.90$\times$R$_0$ to 2.00$\times$R$_0$ 
(R$_0$ is the ground-state separation) are considered.
The interaction types covered by this benchmark set 
include hydrogen bonding, dispersion interactions and other 
interaction types such as X-H$\cdots\pi$ (X=C,N,O) interactions,
as well as nonspecific interactions of polar molecules.
The list of all 66 dimers is presented in Tab.S2 in the
Supporting Information\cite{SuppMat}.  
Minimization of error in the interaction energy of the S66$\times$8 set 
with respect to high-level reference data has been used
by Ambrosetti et al.~\cite{Ambrosetti:14}
to optimize the value of free parameter  of the MBD@rsSCS
method  ($\beta$) and we have followed the same strategy to determine $\beta$ for 
the MBD@rsSCS/FI method. 

The optimal value $\beta$=0.83 found for
MBD@rsSCS/FI is identical 
to that used in the model
based on neutral reference atoms. 
As shown in Tab.~\ref{tab:stat1}, the method based on fractional
ions yields results that are consistently better for all
interaction types covered by this benchmark set
than those obtained by the MBD@rsSCS.  The difference is 
particularly notable for the dispersion interaction dominated subset where
MAE decreased from 17.5~meV to  15.4~meV. 
It is also encouraging that the MBD@rsSCS/FI method improves 
over the MBD@rsSCS for all separations of monomers 
(see Tab. S3 in Supporting Information\cite{SuppMat}). 
The only exception is the largest distance (R=2.00$\times$R$_0$) 
where the interaction energies are small in most cases 
and where the performance of MBD@rsSCS and MBD@rsSCS/FI methods 
is very similar.  

The set of noncovalent molecular dimers X40 of 
\v{R}ez\'a\v{c} et al.~\cite{Rezac:12} 
was designed to benchmark the quality of theoretical methods 
for description of halogen atoms and covers the interaction
types such as London dispersion, induction, 
dipole-dipole, stacking, halogen bonds, halogen-$\pi$ bonds,
and hydrogen bonds.

As a reference, the interaction energies computed at the 
CCSD(T) level are used. The computed interaction energies 
for all systems are compiled in Tab. S4 in the 
Supporting Information\cite{SuppMat},
the statistics obtained by the two variants of the  
MBD@rsSCS method is compared in Tab.~\ref{tab:stat1}. 		
The use of the model based on fractional ions
leads to improvement for all interaction types 
with exception of the dipole-dipole interaction where 
the results are practically identical to those of the 
original method. The most significant 
improvement is found for stacking interactions (MAE reduced 
from $16.8\%$ to $13.3\%$) and  halogen-$\pi$ bonds
($22.5\%$ (MBD@rsSCS) vs. $19.3\%$ (MBD@rsSCS/FI)).
The decrease of error for the whole set is also quite significant 
(MAE: 13.9 meV vs. 15.5 meV; MARE: $9.0\%$ vs. $10.0\%$).  

\subsubsection{Interaction of H$_2$ with substituted coronenes}\label{sec:coro}
\begin{figure}[h]
\begin{center}
$\begin{array}{c@{\hspace{0.0in}}}
\includegraphics[width=.8\columnwidth]{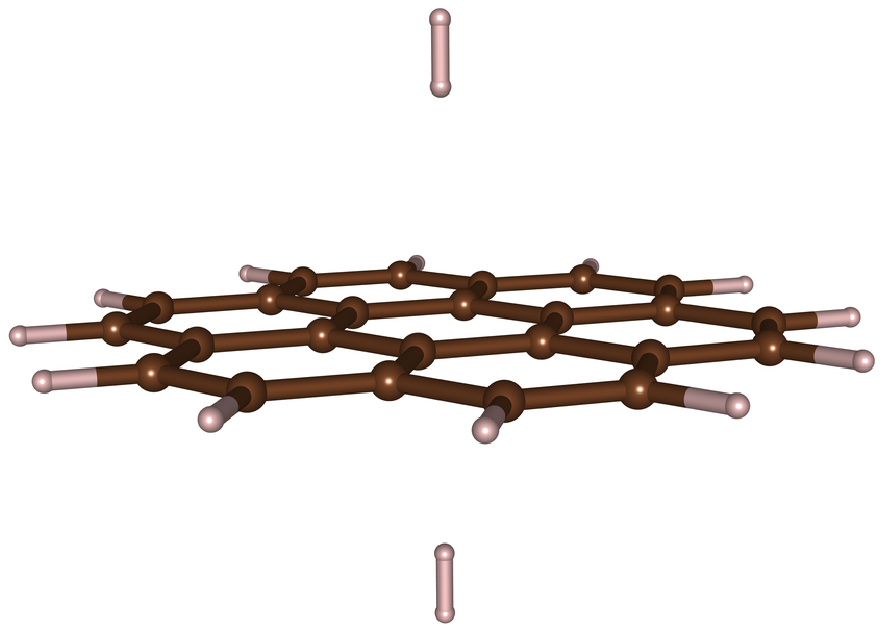}\\
\includegraphics[width=.8\columnwidth]{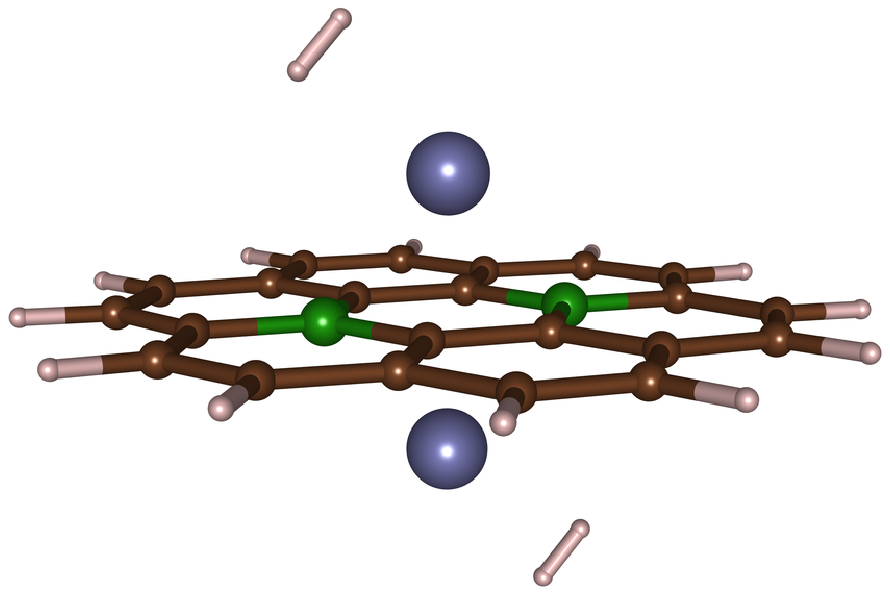} \\
\end{array}$
\end{center}
\caption{Structures of two selected adsorption
complexes from the benchmark set 
of Kocman et al.~\cite{Kocman:15}:
Coronene...2H$_2$ (above), and CoroB$_2$Li$_2$...2H$_2$ (below).
Colour code: C (brown), H (light pink), B (green), Li (blue).
}\label{fig:coro}
\end{figure}

The investigation of hydrogen storage devices represents one of
the important  applications of dispersion corrected DFT methods.
Kocman et al.~\cite{Kocman:15} proposed a set of six adsorption 
complexes consisting of H$_2$ molecules and coronene-derived structures
as a benchmark for methods used in this type of applications. 
The coronene-derived substrate models contain two purely carbonaceous molecules 
(Coronene...2H$_2$ and C-coro...2H$_2$) and four boron
substituted coronene molecules (CoroB$_2$Li$_2$...2H$_2$,
CoroB$_2$...2H$_2$, CoroB$_2$Li$_2$...H$_2$os, 
CoroB$_2$Li$_2$...H$_2$ss) out of which three models contain 
two additional Li$^+$ cations exhibiting relatively strong affinity 
to molecular hydrogen. 
The reference interaction energies determined at the CCSD(T) level
range between 4.7 kJ/mol (Coronene...2H$_2$) to 14.3 kJ/mol
(CoroB$_2$Li$_2$...2H$_2$) and the two extreme cases are shown
in  Fig.~\ref{fig:coro} (all structures are presented in
Figs.~S1-S3 in
the Supporting Information\cite{SuppMat}).

The MBD@rsSCS method based on the neutral atomic reference 
works quite  well for the models that do not contain Li atoms. As shown 
in Tab.~\ref{tab:coro}, the error with respect to the CCSD(T) reference is 
only 1.2 kJ/mol or less in these cases.  The presence of Li atoms,
however, causes
serious numerical issues. In particular, solution of the screening
equation Eq.~\eqref{eqn:SCS} 
yields negative polarizabilities for some atoms and these
can not be handled in the present version of the MBD@rsSCS scheme.

This problem is caused by a very large AIM polarizabilities of the Li atoms 
that is a consequence of fundamental property of the Hirshfeld partitioning
that the atoms-in-molecules are as similar 
to the reference objects (neutral atoms) as possible 
in the information-theoretical 
sense~\cite{Nalewajski:01}. 
Replacing the neutral atomic reference by fractional ions fixes 
this problem: in accord with expectation, the AIM polarizabilities
of the Li in substituted 
coronenes ($\sim$5 au) 
are now closer to polarizability of non-interacting
Li$^+$ (0.2 au~\cite{Gould2016-C6}) than 
to that for free Li$^0$ (164 au~\cite{Gould2016-C6}).
Importantly, the quality of the MBD@rsSCS/FI results for the Li-containing 
models is consistent with that for the purely carbonaceous models 
(see Tab.~\ref{tab:coro}).    

\subsection{Crystalline materials}
\subsubsection{Benchmark set  X23 consisting of molecular crystals}\label{ssec:x23}
In this section we will discuss the performance of different variants of the 
MBD@rsSCS method in optimization of systems from the
X23 set of Reilly et al.~\cite{Reilly:13} 
(based on previous work of Otero-de-la-Roza
and Johnson~\cite{Otero:12})
consisting of 
molecular crystals with cohesion dominated by 
London dispersion and H-bonding interactions.
Full optimization of atomic and lattice degrees of freedom
has been performed and cohesive energies were computed for the 
final geometries. \TG{Simulation}{Computational}
details and additional results
are presented in the Supporting Information\cite{SuppMat}. 

In our previous work~\cite{Bucko:16} we have shown that the neutral atom
based 
model provides already a very good agreement with the reference
energetic and structural data. 
The replacement of neutral atoms by fractional ions
in the polarizability model leads to a 
modest improvement in energetics (MARE$=7.4\%$ (MBD@rsSCS) vs. $6.9\%$ 
MBD@rsSCS/FI) and almost negligible 
variation in predicted lattice structure, see Tab.~\ref{tab:stat2}
and the Supporting Information\cite{SuppMat} (Tabs.S6 and S7).

\subsubsection{Ionic crystals}
Crystalline sodium chloride (space-group $Fm\bar{3}m$) 
represents highly ionic material in which 
the Born effective charges of Na and Cl ions (Z$^*$=$\pm$1.1 e) 
are close to the formal charges $\pm$1 e.~\cite{Bucko:14} 
In our previous work~\cite{Bucko:14} we have shown that
the Hirshfeld partitioning scheme based on neutral atoms strongly 
underestimates the ionic charges (q$^{AIM}$=$\pm$0.2 e). Consequently, and 
in contrast to expectation,
the predicted properties of Na and Cl embedded in the interacting 
systems are more similar to those of neutral atoms than to those 
of ions.

In this case, the pairwise Tkatchenko-Scheffler method 
based on neutral atoms~\cite{Tkatchenko:09} leads to 
a strong overestimation of interaction between atoms in NaCl 
and a strong underestimation of the lattice constant 
[5.34 {\AA} (TS) vs. 5.57 {\AA} (exp.)]\cite{Bucko:14}. 
The consequences for the  MBD@rsSCS scheme based on the 
same polarizability model are 
even more serious: the method suffers from numerical 
issues (already discussed in Sec.~\ref{sec:coro}) that prevent its 
application to systems of this type. 
The use of fractional ions in the MBD@rsSCS/FI scheme 
fixes this problem and the computed 
lattice constant and bulk modulus are in very good 
agreement with experiment, see Tab.~\ref{tab:ionic}.

Magnesium oxide (MgO) is another example of ionic material which we 
will consider here. The crystalline MgO is isostructural to NaCl
and this material is frequently used as a catalyst or adsorbent 
in many applications in chemistry\cite{book:Spivey}.
The standard MBD@rsSCS fails when used to optimze MgO,
while the MBD@rsSCS/FI method yields 
a lattice constant and bulk modulus that are in good agreement 
with experimental data\cite{Sun:11}.
\TG{}{Similar quality results are found for LiF.}

\TG{}{In all three tested ionic solids PBE is accurate, yielding
lattice constants and bulk moduli only slightly worse
than MBD@rsSCS/FI.
However, when considering interactions with surfaces, such as
adsorption or surface reactions\cite{book:Spivey}, it is important
to treat the whole system self-consistently using a method with
vdW corrections. Unlike MBD@rsSCS, MBD@rsSCS/FI is thus accurate
enough to use for surface studies.}

\begin{figure}[h]
\begin{center}
$\begin{array}{c@{\hspace{0.0in}}}
\includegraphics[width=.8\columnwidth]{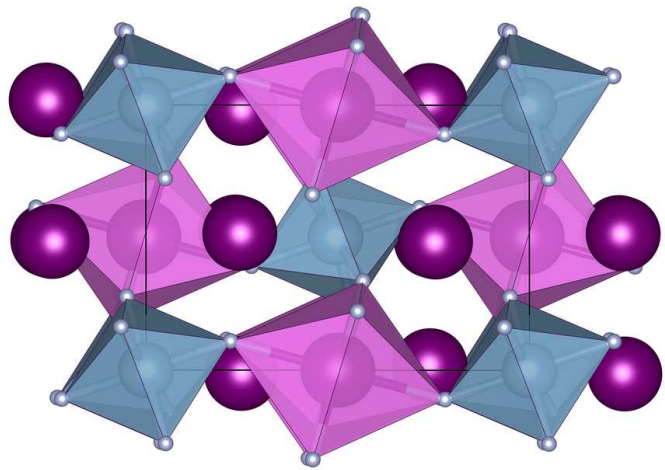}\\
\end{array}$
\end{center}
\caption{Structure of $\alpha$-cryolite. The centers of octahedra are occupied by
Na (violet) or Al (blue) atoms, while the F atoms (small blue spheres) 
sit on their corners. The large 
spheres occupying vacancies between the octahedra are sodium cations. 
The black solid line indicates the monoclinic primitive cell used in calculations. 
}\label{fig:cryolite}
\end{figure}

Next we shall consider mineral cryolite (Na$_3$AlF$_6$), 
which is an example of ionic system with a complex structure.
The low temperature phase $\alpha$ that we will discuss here 
has a symmetry $P2_1/n$ (monoclinic)
and consists of distorted  octahedra AlF$_6$ and NaF$_6$
linked via common 
corners occupied by the fluorine atoms, see Fig.~\ref{fig:cryolite}.
The vacancies between these octahedra are occupied by 
cations Na$^+$.

A recent theoretical study~\cite{Bucko:16b} showed that the semilocal 
DFT (PBE) tends to overestimate the cell volume by $\sim~3\%$ and the error 
even increases in the high-temperature phase $\beta$ and in the
liquid phase.
The missing dispersion interactions have been suggested 
as a possible reason for this volume overestimation. 
The lattice geometry determined using the MBD@rsSCS/FI 
($a=5.39$ {\AA}, $b=5.61$ {\AA}, $c=7.78$ {\AA}, 
$\beta=90.3$\degree)
is close to the experimental geometry for T=0K~\cite{Yang:93} 
($a=5.38$ {\AA}, $b=5.58$ {\AA}, 
$c=7.69$ {\AA}, $\beta=90.3$\degree) and the error in
the computed cell volume is $<1\%$. 
By contrast, the attempted MBD@rsSCS calculation failed for the same reason 
as in the case of Li-substituted coronenes and NaCl.

\subsubsection{Dichalcogenides}\label{ssec:dichalco}

In our previous work, we have shown that the conventional
MBD@rsSCS scheme works very well in description of the structure
and energetics of graphene bilayers and graphite~\cite{Bucko:16}. 
Here we focus on the properties of bilayers and bulk systems 
of dichalcogenides (MoS$_2$, MoSe$_2$,
WS$_2$, and WSe$_2$), which can be considered as van der Waals 
systems with a partially ionic character. 

For the bilayers of dichalcogenides discussed here, the two most 
favorable stacking patterns AA' and AB~\cite{He:14} are considered
and the RPA results of He et al.~\cite{He:14} are used as a reference.
In the calculations, the interlayer distance was varied while 
the geometry of individual layers was fixed at the value 
determined experimentally for the bulk systems ~\cite{Bjorkman:12}. 

Because of the problems with incorrect eigenvalue spectrum of 
the $\ten{A}_{\LR}(\omega)\ten{T}_{\LR}$ matrix (see Sec.~\ref{ssec:Rescale}),
the original variant of the MBD@rsSCS method can not be applied
directly  to the systems discussed here, hence we discuss only results 
obtained using MBD@rsSCS/FI method in which this problem
is overcome, as described in Sec.~\ref{ssec:Rescale}.

\begin{figure}[h]
\begin{center}
$\begin{array}{c@{\hspace{0.0in}}}
\includegraphics[width=0.8\columnwidth]{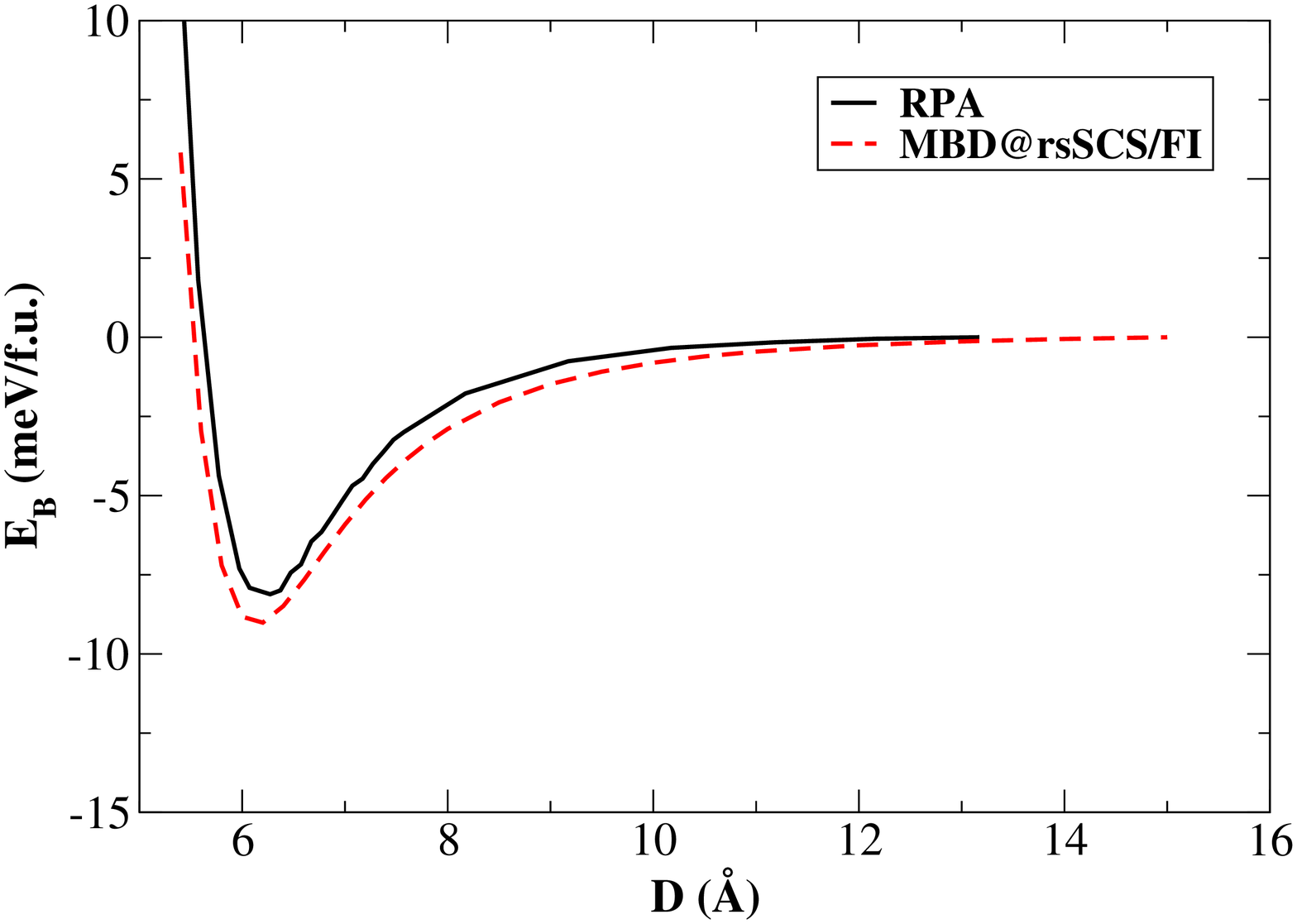} \\
\includegraphics[width=0.8\columnwidth]{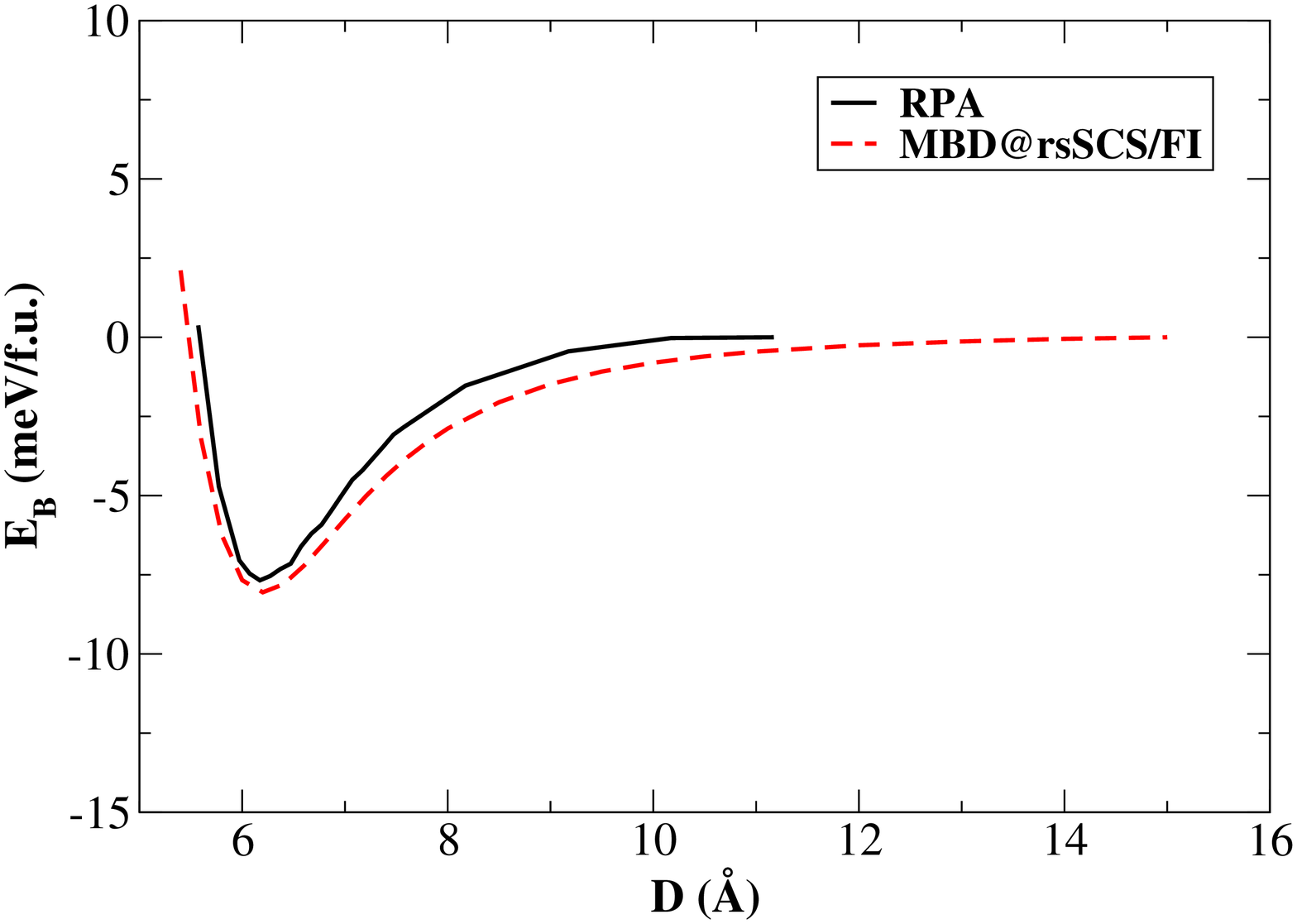} 
\\
\end{array}$
\end{center}
\caption{Binding energy of bilayers of
  MoS$_2$ as a function of interlayer distance (D):
  AA' stacking (top), AB stacking (bottom). The reference RPA
  results are from \rcite{He:14}.
}\label{fig:mos2}
\end{figure}

The computed interaction energy profiles for
the AA'- and AB-stacked MoS$_2$ are compared with the RPA reference 
of Hu et al.~\cite{He:14} in Fig.~\ref{fig:mos2};
the profiles for the other systems considered here 
(MoSe$_2$, WS$_2$, and WSe$_2$) are presented in Figs.~S4-S7
in the Supporting Information\cite{SuppMat}. 
As evident from the comparison of the energy profiles 
and from the numerical values of binding energies and 
equilibrium interlayer separations
(see Tab.~\ref{tab:bilayers}), the MBD@rsSCS/FI
method reproduces the RPA results very well and the improvement 
over the pairwise version of the method (PBE+TS~\cite{Tkatchenko:09})
is remarkable.

The calculations of ground-state structure and binding energies of
bulk dichalcogenides with AA' stacking have been performed
via series of single-point 
energy calculations in which the lattice parameter parallel with 
the stacking direction ($c$) was varied in the interval between 
$\sim$10.7 {\AA} to 44.2 {\AA}. 
In these calculations, all intralayer structural parameters 
and lattice vectors parallel to the layers were fixed at
their experimental values.

The computed results presented in Tab.~\ref{tab:bulklayered}
are qualitatively consistent with our results for bilayers and
 the MBD@rsSCS/FI results are
reasonably close to the RPA reference. 
The standard MBD@rsSCS method has the same numerical issues
(from the polarization catastrophe) as in the case of bilayers.
We thus cannot calculate or report meaningful results from it.

\subsubsection{Graphite fluoride}
\begin{figure}[h]
\begin{center}
$\begin{array}{c@{\hspace{0.0in}}}
\includegraphics[width=.8\columnwidth]{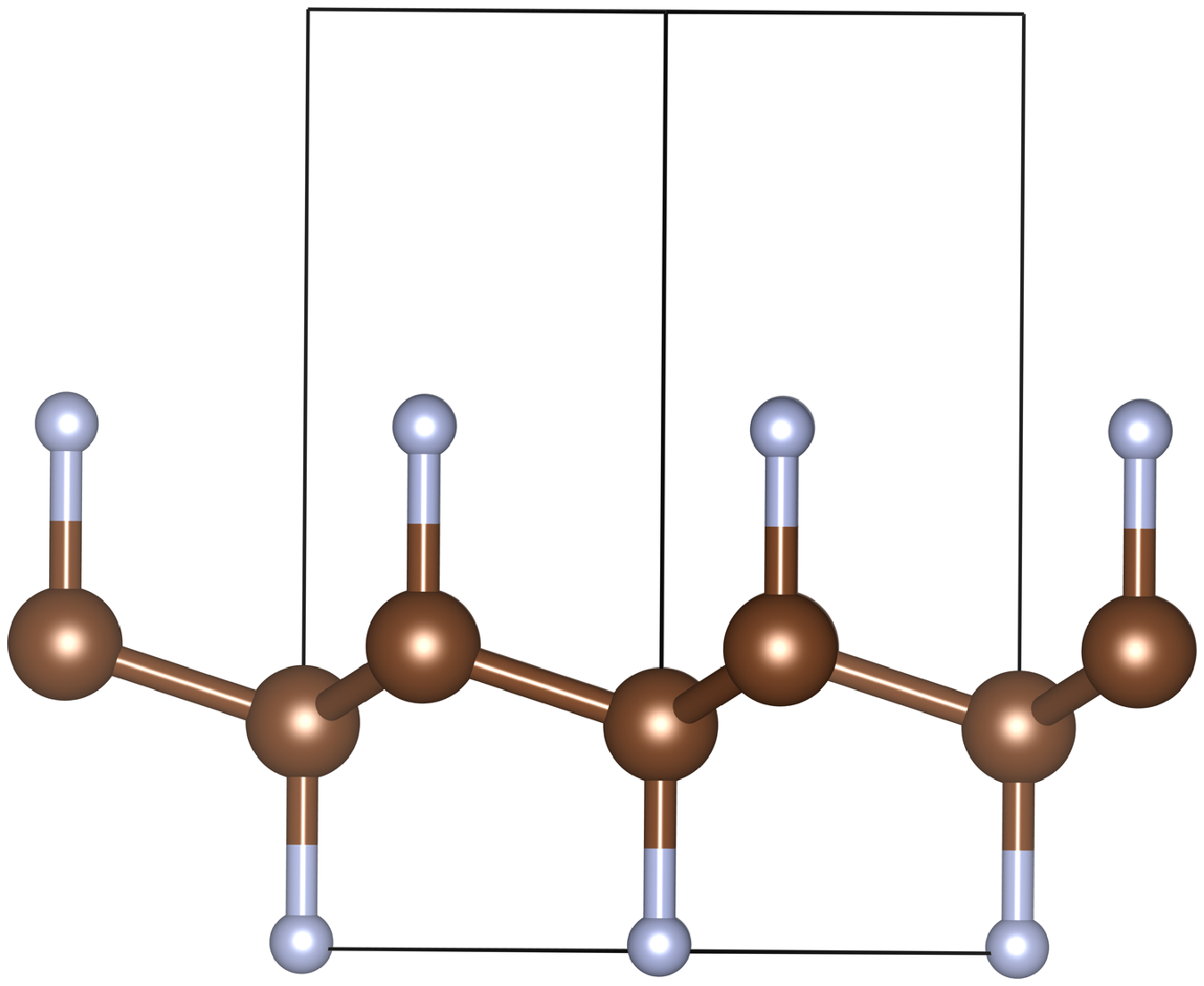} \\
\includegraphics[width=.8\columnwidth]{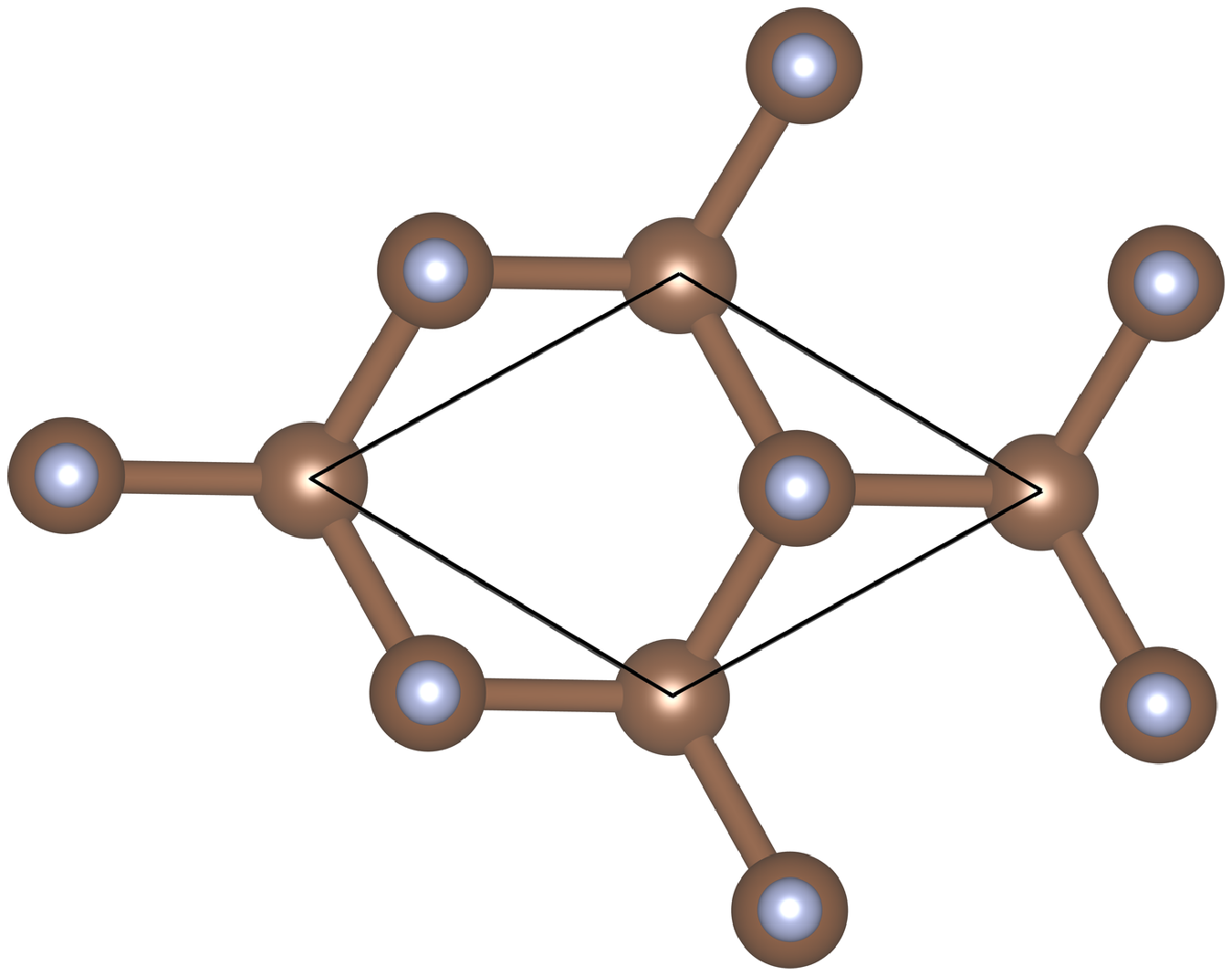} 
\\
\end{array}$
\end{center}
\caption{Side and top view of the structure of the AA stacked graphite fluoride.
}\label{fig:CFstruct}
\end{figure}

In fluoride derivative of  graphite, the fluorine atoms 
are covalently bound to carbon atoms, see Fig.~\ref{fig:CFstruct}. 
In this work we use the MBD@rsSCS 
approach to compute binding energy for the 
structure with the AA stacking of layers which has 
recently been identified as the most stable structural arrangement 
for this material~\cite{Lazar:15}. 
The simulation cell used in calculations contained four atoms belonging 
to a single layer and the interlayer separation was controlled via
the length of the lattice vector normal to the
layer ($c$). Following 
Lazar et al.~\cite{Lazar:15}, atomic positions 
and length of lattice vectors parallel with the layer ($a$=2.582 {\AA}) were 
fixed during the calculation.

\begin{figure}[h]
  \caption{Binding energy of graphite fluoride
  as a function of interlayer distance (D). The reference RPA
	results are from \rcite{Lazar:15}.  
  }\label{fig:CF}
  \includegraphics[width=\columnwidth]{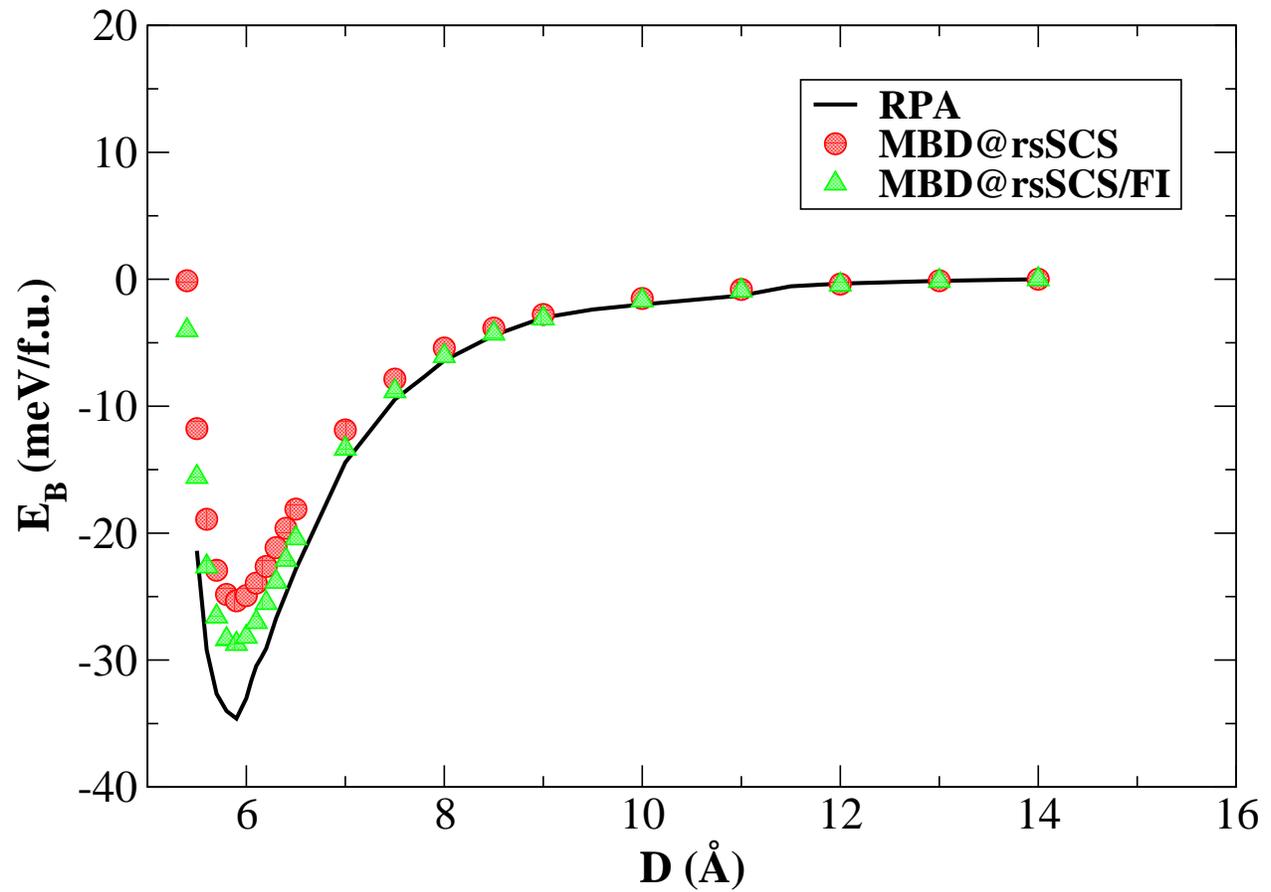}
\end{figure}

The dependence of the binding energy on
the interlayer distance computed using the MBD@rsSCS and MBD@rsSCS/FI methods 
is compared with the RPA reference~\cite{Lazar:15} in Fig.~\ref{fig:CF}. 
The MBD@rsSCS and MBD@rsSCS/FI methods predict a similar ground 
state value for the interlayer 
separation (5.91 {\AA} (MBD@rsSCS), 5.89 {\AA} (MBD@rsSCS/FI))
that are close to the RPA reference (5.88 {\AA}).
The binding energy is somewhat underestimated in both cases 
but the MBD@rsSCS/FI value (28.7 meV/fu) is closer 
to the RPA reference (34.7 meV/fu) than the MBD@rsSCS
result (25.3 meV/fu).

\section{Limitations and perspectives}

\Note{TG}{***}
Before concluding, we discuss a few limitations of our FI approach
which warrant study in future works. Many of these limitations are
shared by other van der Waals functionals.

Firstly, the FI approach neglects quadropolar and
higher order van der Waals terms. This can possibly be remedied by
approximating these terms\cite{Tao2016-Higher} or pre-calculating
them directly in higher level theory as in \rcite{Gould2016-C6}.
However, we note that in layered geometries the higher order
contributions appear to be heavily screened in the near-contact
regime\cite{Dobson2016-LRT}, perhaps explaining the success
of methods which neglect them.

Secondly, the lack of long-range fluctuations in the scheme
means it cannot properly capture
Dobson-C\cite{Dobson2014-IJQC} effects, a property it inherits
from the MBD scheme. This is likely to be important in long-range
forces from metallic systems\cite{Dobson2006,Dobson2009}
and systems with unusual topological
properties\cite{Dobson2014-1}. However, there is some
evidence (e.g.~\cite{Ambrosetti2016}) that some of these effects
might be captured via the coupled fluctuating dipole model's
inclusion of Dobson-B terms, at least in the medium range.

Finally, the eigenvalue remapping scheme is somewhat \emph{ad hoc}.
Further testing of how the polarization catastrophe behaves in
exemplar (non-remapped) systems should shed light on how to
include more physically intuitive corrections. Work in this
direction is being pursued.
\Note{TG}{***}

\section{Conclusions}

In this work we introduced a scheme (MBD@rsSCS/FI)
for calculating polarizabilities and dispersion forces
in electronic systems. Our approach uses the properties
of \emph{fractional ions} (FIs) in a point-dipole dispersion
approximation based on the MBD approach.
Key equations describing our approach are
given in Eqs.~\ref{eqn:alphaTS2}, \ref{eqn:alpha}
and \ref{eqn:EReqn}, and related discussion.

Our fractional ionic approach vastly improves results for ionic
systems in bulk and layered materials, compared to related
approaches\cite{Tkatchenko2009,Tkatchenko2012,DiStasio2012,Bucko2014-Iterative}.
It is particularly
successful in transition-metal dichalcogenides like MoS$_2$,
and in interactions of H$_2$ with modified coronenes,
systems of nanotechnological interest.
It marginally outperforms similar schemes in most
of the non-ionic system sets tested, and is of only
marginally lower accuracy in the remainder.

Our approach is thus a suitable alternative to existing schemes
in systems where they work, while offering substantial improvements
in ionic systems where they do not.
Therefore we strongly advocate in favour of the
PBE+MBD@SCS/FI method in routine applications
\TG{}{(using keyword {\tt IVDW=263} in VASP)}.

\begin{acknowledgement}
T.G. received computing support from the
Griffith University Gowonda HPC Cluster
and a Griffith University International Travel Fellowship.
Part of this work has been supported by the VASP project. 
T.B. is grateful to University of Lorraine for invited professorships
during the academic year 2015-16 and he
acknowledges support from Project
643 No. APVV-15-0105. Calculations were performed
using
computational resources of University of Vienna, and
supercomputing infrastructure of Computing Center of the 
Slovak Academy of Sciences acquired in projects ITMS 26230120002 and
26210120002 supported by the
Research and Development Operational Program funded by the ERDF.
J.G.A. is thankful for the partial support of this research by
the European Union and the State of Hungary, co-financed by the
European Social Fund in the framework of
T{\'A}MOP 4.2.4. A/2-11-1-2012-0001 National Excellence Program.
S. L. acknowledges HPC resources from GENCI-CCRT/CINES
(Grant x2017-085106).
\end{acknowledgement}

\begin{suppinfo}
We include with this manuscript supporting information including:
\begin{itemize}
\item[] A PDF document that includes:
  further theoretical details of methods;
  values of computed atoms-in-molecule polarizabilities;
  and numerical results for S22 and X40 benchmark sets.
\end{itemize}
\end{suppinfo}

\providecommand{\latin}[1]{#1}
\providecommand*\mcitethebibliography{\thebibliography}
\csname @ifundefined\endcsname{endmcitethebibliography}
  {\let\endmcitethebibliography\endthebibliography}{}

\newpage
\clearpage
\begin{table}[h!]
\caption{
Summary of simulation parameters used in this study. 
See Tab.S5 in the Supporting Information\cite{SuppMat}
for details on setting used 
in X23 set calculations. 
}
\label{tab:setting}
\begin{tabular}{lcc}
\hline
 System & $\vec{k}$-point mesh & $E_{\text{cut}}$ (eV) \\
\hline
S66$\times$8& $1\times 1 \times 1$& 1000\\
X40& $1\times 1 \times 1$& 1000\\
H$_2$+coronenes&  $1\times 1 \times 1$& 1000\\
NaCl& $16\times 16 \times 16$& 1000\\
MgO& $16\times 16 \times 16$& 1000\\
cryolite& $8\times 8 \times 6$& 1000\\
bilayers of dichalcogenides& $16\times 16 \times 1$& 1000\\
bulk dichalcogenides& $12\times 12 \times 3$& 1500\\
fluorographite& $20\times 20 \times 8$& 1000\\
\hline 
\end{tabular}
\end{table}

\begin{table*}[h]
\footnotesize
\caption{Mean absolute error (MAE) and mean absolute relative 
error (MARE) for interaction energies computed for
the S66$\times$8~\cite{Rezac:11} and the X40~\cite{Rezac:12} benchmark sets computed 
using  the MBD@rsSCS and MBD@rsSCS/FI methods. 
The CCSD(T) results from \rcites{Rezac:11,Rezac:12} have been
used as a reference.
Statistics for  
subsets dimers differing in the nature of dominating interaction is also 
listed (the numbers in parentheses indicate the number of dimers in the 
given subset).
}\label{tab:stat1}
\begin{tabular}{lrcccc}
\hline
set& subset  & \multicolumn{2}{c}{MBD@rsSCS}& 
\multicolumn{2}{c}{MBD@rsSCS/FI}\\
   & & MAE (meV)& MARE ($\%$)&  
	MAE (meV)& MARE ($\%$)\\
\hline
S66$\times$8&  all (66$\times$8)& 13.6& 9.7&12.3& 9.0\\
          & H-bond (23$\times$8)& 19.1& 8.3&17.8& 7.8\\        		    
      & dispersion (23$\times$8)& 17.5& 6.4&15.4& 5.7\\
	     		 & other (20$\times$8)&  8.9& 6.6&8.1& 6.2\\            
\hline
X40& all (40)& 15.5& 10.0& 13.9& 9.0\\ 
& dispersion (4)& 3.9& 16.4& 3.6& 15.9\\
   &  induction (4)& 3.0& 8.5& 2.5& 7.3\\
 &dipole-dipole (2)& 8.8& 13.6& 8.9& 13.7\\
	  &  stacking (2)& 38.4& 16.8&30.2& 13.3\\
  &halogen-bond (14)& 5.5& 5.0& 4.7& 4.3\\
& halogen-$\pi$ (4)& 26.4& 22.5&  23.3& 19.3\\
&        H-bond (10)& 31.5& 8.0&  29.3& 7.5\\ 
\hline						
\end{tabular}
\end{table*}

\begin{table*}[h]
\footnotesize
\caption{Interaction energy per a H$_2$ molecule (kJ/mol) 
for set of complexes of carbonaceous and substituted coronenes. 
The CCSD(T) results are from Kocman et al.~\cite{Kocman:15} 
}\label{tab:coro}
\begin{tabular}{lccc}
\hline
  & CCSD(T)& MBD@rsSCS&  MBD@rsSCS/FI\\
\hline
        Coronene...2H$_2$&  $-$4.7& $-$4.5&    $-$4.6\\
 CoroB$_2$Li$_2$...2H$_2$& $-$14.3&  crash&  $-$15.9\\
       CoroB$_2$...2H$_2$&  $-$4.9& $-$3.7&    $-$3.7\\
CoroB$_2$Li$_2$...H$_2$ss& $-$10.4&  crash&  $-$12.9\\
CoroB$_2$Li$_2$...H$_2$os&  $-$5.0&  crash&    $-$5.6\\
          C-coro...2H$_2$&  $-$5.5& $-$4.9&    $-$5.0\\  		
\hline
\end{tabular}
\end{table*}

\begin{table*}[h]
\footnotesize
\caption{Statistics for interaction energy and cell volume 
for molecular crystals from the benchmark set  X23 computed 
using the MBD@rsSCS and MBD@rsSCS/FI methods. 
The reference values are from \rcite{Reilly:13}. Details
on energetics and relaxed structrures of individual systems are 
given in the Supporting Information\cite{SuppMat}.
}\label{tab:stat2}
\begin{tabular}{lcc}
\hline
  & MBD@rsSCS& MBD@rsSCS/FI\\				
\hline							
      MAE(E) (kJ/mol)& 5.7&  5.2\\
		   MARE(E) ($\%$)& 7.4&  6.9\\
MAE(V) ({\AA}$^3$/fu)& 1.9 & 2.0\\
       MARE(V) ($\%$)& 1.9 & 2.0 \\
\hline
\end{tabular}
\end{table*}

\begin{table*}[h]
\footnotesize
\caption{Static lattice constants and bulk moduli for
  crystalline NaCl, MgO and LiF. The zero-temperature experimental
  values corrected for zero-point phonon effects 
  are from \rcite{Sun:11}.
}\label{tab:ionic}
\begin{tabular}{lcccccc}
  \hline
  & \multicolumn{2}{c}{NaCl}& \multicolumn{2}{c}{MgO}&
  \multicolumn{2}{c}{LiF}\\
  & $a$ ({\AA})& $B_0$ (GPa)& $a$ ({\AA})& $B_0$ (GPa)&
  $a$ ({\AA})& $B_0$ (GPa)\\
  \hline
  Exp.& 5.57 & 28 & 4.19 & 170 & 3.97 & 76\\
  PBE & 5.55 & 25 & 4.25 & 152 & 4.07 & 67\\
  MBD@rsSCS& \multicolumn{2}{c}{crash}& \multicolumn{2}{c}{crash}
  & \multicolumn{2}{c}{crash}\\
  MBD@rsSCS/FI& 5.62 & 26 & 4.20 & 166 & 4.05 & 68 \\
  \hline
\end{tabular}
\end{table*}

\begin{table*}[h]
  \footnotesize
  \caption{Interlayer distances $d_0$,  and interlayer binding
    energies $E_b$ for AA' and AB stacking variants
    of bilayers of transition-metal dichalcogenides from RPA and dispersion corrected 
		DFT calculations. The RPA and pairwise TS results are from He et
    al. (\rcite{He:14}) and Bu\v{c}ko et al. (\rcite{Bucko:14}), 
		respectively.}
  \label{tab:bilayers}
\scriptsize
    \begin{tabular}{{l}*{8}{c}}
\hline		
      & \multicolumn{2}{c}{RPA}
			& \multicolumn{2}{c}{TS}
			& \multicolumn{2}{c}{MBD@rsSCS} 
			& \multicolumn{2}{c}{MBD@rsSCS/FI}\\
 System(stacking)   &  $d_0$ ({\AA})  & $E_b$ (meV/fu) & $d_0$ ({\AA})  & $E_b$ (meV/fu)
& $d_0$ ({\AA})  & $E_b$ (meV/fu)
& $d_0$ ({\AA})  & $E_b$ (meV/fu)\\
\hline
 MoS$_2$(AA')& 6.27& 81.2& 6.08& 149.2& \multicolumn{2}{c}{crash}&   6.15&  90.3\\
  MoS$_2$(AB)& 6.17& 76.7& 6.08& 140.4& \multicolumn{2}{c}{crash}&   6.22&  80.6\\
\hline
MoSe$_2$(AA')& 6.48& 88.4& 6.44& 148.7& \multicolumn{2}{c}{crash}&   6.42& 110.2\\
 MoSe$_2$(AB)& 6.47& 85.1& 6.55& 135.6& \multicolumn{2}{c}{crash}&   6.61& 88.7\\
\hline
  WS$_2$(AA')& 6.24& 82.9& 6.18& 134.2& \multicolumn{2}{c}{crash}&   6.26& 76.7\\
   WS$_2$(AB)& 6.24& 74.3& 6.28& 124.7& \multicolumn{2}{c}{crash}&   6.37& 67.5\\
\hline
 WSe$_2$(AA')& 6.50& 89.9& 6.52& 135.8& \multicolumn{2}{c}{crash}&   6.54& 93.1\\ 
  WSe$_2$(AB)& 6.54& 84.2& 6.73& 123.5& \multicolumn{2}{c}{crash}&   6.75& 76.0\\
\hline
MAE& -& -& 0.09& 53.7& \multicolumn{2}{c}{-}&   0.10& 7.8\\
\hline
\end{tabular}
\end{table*}

\begin{table*}[h]
\footnotesize
\caption{Cell geometry and interlayer binding energy $E_{B}$ for 
selected dichalcogenides. Experimental structures and the energies
computed at the RPA level~\cite{Bjorkman:12} are used as a reference.
The pairwise TS results are from \rcite{Bucko:14}.
}\label{tab:bulklayered}
\begin{tabular}{llccc}
\hline
Compounds & Method &          $a$ ({\AA}) &  $c$ ({\AA})& $E_{B}$ (meV/{\AA}$^2$) \\
\hline
MoS$_2$       & Expt./RPA& 3.16& 12.29& 20.5\\
              &       TS &     & 12.09&  37.9\\ 
              & MBD@rsSCS& -& \multicolumn{2}{c}{crash}\\
  &MBD@rsSCS/FI& -& 12.28& 20.7\\
\hline 
      MoSe$_2$& Expt./RPA& 3.29& 12.90& 19.6\\
			        &       TS &     & 12.78&  34.0\\
			        & MBD@rsSCS& -& \multicolumn{2}{c}{crash}\\
  &MBD@rsSCS/FI& -& 12.78& 20.6\\
\hline
WS$_2$         & exp./RPA& 3.15& 12.32& 20.2\\
               &       TS &     & 12.22&  34.2\\
               & MBD@rsSCS& -& \multicolumn{2}{c}{crash}\\
  &MBD@rsSCS/FI& -& 12.46& 18.3\\
\hline
WSe$_2$       & Expt./RPA& 3.28& 12.96& 20.0\\
              &       TS &     & 12.97&  31.3\\
              & MBD@rsSCS& -& \multicolumn{2}{c}{crash}\\
  &MBD@rsSCS/FI& -& 13.01& 19.8\\
\hline
\end{tabular}
\end{table*}

\newpage
\clearpage

\end{document}